\documentstyle[preprint,aps]{revtex}
\tighten
\begin{document}
\draft
\title{\bf
 Shell  corrections  for finite depth potentials:\\
Particle  continuum effects
 }
\author  {T. Vertse,$^{1-3}$,
	   A.T. Kruppa,$^1$,
	   R.J. Liotta,$^3$
	   W. Nazarewicz,$^{4-6}$
	   N. Sandulescu,$^{3,7}$, \\
           and T.R. Werner$^6$
	   }
\address {
	  $^1$Institute of Nuclear Research of the Hungarian
	  Academy of Sciences \\
	  P. O. Box 51, H-4001, Debrecen, Hungary \\[1mm]
$^2$Joint Institute for Heavy Ion Research\\
   Oak Ridge National Laboratory,
   P.O. Box 2008, Oak Ridge,   Tennessee 37831 \\[1mm]
	  $^3$Royal Institute of Technology, Physics Department
	  Frescati \\
	  Frescativagen 24, S-10405, Stockholm, Sweden \\[1mm]
	  $^4$Department of Physics and Astronomy, University
	  of Tennessee \\
Knoxville  Tennessee 37996 \\[1mm]
$^5$Physics Division, Oak Ridge National Laboratory \\
   P.O. Box 2008, Oak Ridge,   Tennessee 37831, \\[1mm]
	  $^6$Institute of Theoretical Physics, Warsaw University,
	  ul. Ho\.za 69, PL-00681, Warsaw, Poland\\[1mm]
	  $^7$Institute of Atomic Physics, Box MG-6, Bucharest,
	  Romania
}

\maketitle
\begin{abstract}

Shell corrections of finite, spherical, one-body potentials are
analyzed using a smoothing procedure which properly accounts
for the contribution from the particle continuum, i.e.,
unbound states.  Since the  plateau
condition for the smoothed single-particle energy seldom holds,
a  new recipe is suggested for the definition of the
shell correction.  The generalized Strutinsky smoothing
procedure is compared with the results of the semi-classical
Wigner-Kirkwood  expansion. A good agreement has been found for
weakly bound nuclei in the vicinity of the proton drip line.
However, some deviations remain for extremely neutron-rich
systems due to the  pathological behavior of the
semi-classical  level density around the particle threshold.

\end{abstract}


\narrowtext

\section{Introduction}

Positive-energy  eigenstates of the average single-particle
potential are very important for
the description
of nuclei
close to the particle  drip lines where the Fermi level
approaches zero (see Ref.~\cite{[Dob97a]} and references
quoted therein), and
in analysis of highly excited nuclear modes such as giant
resonances \cite{San97,[Ver95]}.
With the advent of radioactive nuclear beams,
of particular interest are masses of weakly bound nuclei with
 extreme
$N/Z$ ratios. For these nuclei, important for both nuclear structure and
astrophysics, special  care should be taken when  dealing
with the particle continuum,  which  strongly influences many
nuclear properties, including global ones (e.g., masses, radii, shapes) as well
as nuclear dynamics (i.e., excitation modes).

In a  previous paper \cite{Naz94}, a
macroscopic-microscopic  method was
applied
to nuclei far from the beta stability line.
It has been demonstrated that
 the systematic error
in binding energies,
due to the particle continuum,
can be as large as several MeV at the neutron drip line;
 hence it
can
seriously affect theoretical
mass predictions for nuclei far from stability.
This error depends weakly on deformation, thus
suggesting a possibility of re-normalizing potential energy surfaces
at the spherical shape.

In this paper, the effect of the particle continuum on the shell
correction (the quantal contribution to the total energy
in the macroscopic-microscopic approach)
is investigated by solving the Schr\"odinger equation in the
complex energy plane. The new  procedure allows for the direct treatment
of both narrow
resonances and the smooth continuum background when calculating
the single-particle level density.

The paper is organized as follows. Section~\ref{shellcorr} contains
a  brief review of the shell-correction method in terms of the single-particle
level density. The semi-classical approach is discussed in
Sec.~\ref{semiclassical}.
Section~\ref{newtreatment} describes the modified Strutinsky re-normalization
procedure which takes care of the continuum effects.
The results of the calculations are contained in Sec.~\ref{results}.
Finally, conclusions are given  in Sec.~\ref{conclusions}. The threshold
behavior of the semi-classical level density is discussed
 in  Appendix~\ref{appA}.

\section{Shell Correction and Single-Particle Level Density}\label{shellcorr}

In the standard macroscopic-microscopic approach \cite{Str67,Str68,Bol72,Mo81},
 the shell correction
\begin{equation}\label{e1}
\delta E = E_{\rm s.p.} - \tilde{E}_{\rm s.p.}
\end{equation}
 is the difference between  the
total single-particle
 energy  $E_{\rm s.p.}$,
\begin{equation}\label{shell}
E_{\rm s.p.} =
\sum_{i-{\rm occ}}\varepsilon_i\ ,
\end{equation}
and  the  smooth single-particle  energy  $\tilde{E}_{\rm s.p.}$.
The shell correction represents the fluctuating part of the binding
energy resulting from the single-particle shell structure.

For the sake of simplicity,  we shall restrict our discussion
to spherically symmetric nuclei and assume that the single-nucleon
energy spectrum is  that of a
one-body  Hamiltonian $\hat{H}=T+V$ with a finite, local, and spherically
symmetric potential
$V(r)$. Since the spectrum of $H$ contains both bound
($\varepsilon_i<0$) and unbound
 ($\varepsilon>0$) eigenvalues,
the single-particle level density
is the sum of the discrete and continuum
contributions \cite{Uhl36,Bet37,Li70,Ro72,Shl92}
\begin{equation}\label{e3}
g(\varepsilon) = \sum_i ( 2 j_i + 1 )\delta(\varepsilon-\varepsilon_i) + g_{\rm c}(\varepsilon).
\end{equation}
The  continuum part, $g_{\rm c}(\varepsilon)$, is defined in terms of the
scattering phase shifts $\delta_{lj}(\varepsilon)$
\begin{equation}\label{e4}
g_{\rm c}(\varepsilon)={1\over \pi}
\sum_{l,j}(2j+1)\frac{d\delta_{lj}(\varepsilon)}{d\varepsilon}.
\end{equation}

In the shell-correction method
\cite{Str67,Str68} the smooth energy $\tilde{E}_{\rm s.p.}$
is calculated by employing  the
smooth  level density $\tilde{g}(\varepsilon)$ obtained from $g(\varepsilon)$
by folding with a smoothing function
$f(x)$:
\begin{equation}\label{ggsmooth}
\tilde{g}(\varepsilon) = {1\over\gamma}
\int_{-\infty}^{+\infty}d\varepsilon'~ g(\varepsilon')
f\left(\frac{\varepsilon'-\varepsilon}{\gamma}\right).
\end{equation}

The folding function $f(x)$
can be written as a product of a
weighting  function  and a curvature
correction
polynomial of the order $p$=$2M$ \cite{Str68}.
The smoothing procedure should be unambiguous, i.e.,
the averaging should leave
the smooth part of the level density  untouched
(the so-called
self-consistency condition for the Strutinsky smoothing).
This  defines  a curvature
correction
polynomial for
any specific choice of weighting  function.
In this study,  a Gaussian
weighting  function, $\frac{1}{\sqrt{\pi}}\exp(-x^2)$
has been used.
The corresponding curvature function is an associated Laguerre
polynomial $L_M^{1/2}(x^2)$. This choice guarantees
the self-consistency condition for $\tilde{g}(\varepsilon)$
if the smooth level density behaves as a polynomial of degree
$2M+1$ or lower in $\varepsilon$.

The smoothed level density (\ref{ggsmooth}) defines both the
 smooth single-particle  energy
\begin{equation}\label{Esmoth}
\tilde{E}_{\rm s.p.}=\int_{-\infty}^{\tilde\lambda}
\varepsilon\, \tilde{g}(\varepsilon) d\varepsilon,
\end{equation}
and  the smoothed Fermi level $\tilde{\lambda}$. The latter is
obtained  from
the particle number equation:
\begin{equation}\label{ltilde}
N = \int_{-\infty}^{\tilde{\lambda}} \tilde{g}(\varepsilon)d\varepsilon,
\end{equation}
where $N$ is the  total number of particles (i.e. protons or neutrons).
The smoothing range $\gamma$  should be
greater  than  the average energy distance between
neighboring major shells,  $\hbar\Omega$$\approx$41/$A^{1/3}$\, MeV
\cite{[Boh69]}.

Since the result of the smoothing procedure should not depend on the actual
form of the smoothing function, in particular on the smoothing range $\gamma$
and the order $p$ of curvature correction,
the smooth energy should satisfy the  so-called {\em plateau condition}:
\begin{equation}\label{plateau}
\frac{\partial\tilde{E}_{\rm s.p.}}{\partial\gamma}=0\ \ ,
\frac{\partial\tilde{E}_{\rm s.p.}}{\partial p}=0.
\end{equation}
For  infinite potentials such as a
harmonic oscillator or a deformed Nilsson potential,
one can always find an interval
of the smoothing parameters $\gamma$ and $p$ in which the
 smooth energy, hence the  shell correction, is practically independent of the
values  $\gamma$ and $p$ \cite{Ro72,Bra73}.
For  finite-depth  potentials, additional complications
 arise due to  the presence of
 the continuum
contribution, Eq.~(\ref{e4}) (see discussion in Refs.~\cite{Naz94,San97}
and references quoted therein).
In most calculations applying the shell-correction
approach,
 the continuum is treated approximately
 by using the {\em quasi-bound states}, i.e., the states
resulting from the diagonalization of a finite potential in a large
harmonic oscillator basis \cite{Bol72}. However, for
light and  weakly bound nuclei, the plateau condition (\ref{plateau})
usually does not hold  \cite{Naz94}.

\section{Semi-classical Treatment of Shell Correction}\label{semiclassical}

A possible alternative to Strutinsky's smoothing procedure
is the semi-classical averaging based on the
Wigner-Kirkwood expansion
\cite{Bra73,Bha71,Jen73,Jen75,Jen75a,Jen75b,Bra97}.
In Ref.~\cite{Jen73}  the equivalence between the Strutinsky
approach and
the Wigner-Kirkwood (WK) expansion has been demonstrated.
It is important to note that
this proof assumed that
in the Strutinsky approach the plateau  condition could be  fulfilled.

In the WK
expansion, the diagonal part of the the Bloch density 
\cite{Bra97,Bh77} is
\begin{equation}\label{denmat}
C(\bbox{r},\beta)={1\over 4\pi^{3/2}\beta ^{3/2}}\left (
{2M\over \hbar ^2}\right )^{3/2} 
e^{-\beta V(\bbox{r})} 
\left\{ 1- {{\hbar ^2\beta^2}\over {12M}} 
\left[
\nabla^2V(\bbox{r}) 
- \frac{\beta}{2} (\nabla V(\bbox{r}))^2
\right]+\ldots \right\}.
\end{equation}
The spatial  density is obtained 
by using the inverse Laplace transform
\begin{equation}\label{partden}
\rho(\bbox{r},\varepsilon)={\cal L}^{-1}_{\beta\rightarrow\varepsilon}
\left ( {C(\bbox{r},\beta)\over\beta}\right ),
\end{equation}
and the particle number integral  is given by
\begin{equation}\label{totN}
N(\varepsilon)=\int \rho(\bbox{r},\varepsilon)\ d^3r.
\end{equation}

Keeping only the two leading terms in the curly bracket
in Eq.~(\ref{denmat}), the WK particle number equation
can be expressed explicitly in terms 
of the single-particle potential
\begin{equation}\label{Npart1}
N_{sc}(\varepsilon) =
\frac{4}{3\pi}\left(\frac{2M}{\hbar^2}\right)^{3\over 2}
\int^{r_{\rm sc}(\varepsilon)}  r^2\left\{(\varepsilon-V)^{3\over 2}
- \frac{\hbar^2}{32M}\frac{\nabla^2 V}{(\varepsilon-V)^{1\over
2}}\right\} dr.
\end{equation}

The integral in Eq.~(\ref{Npart1}) is cut off at the
classical turning point, $r_{\rm sc}(\varepsilon)$,
defined by the relation
$V(r_{\rm sc})=\varepsilon$. (For the inclusion of the
spin-orbit term see Ref.~\cite{Jen75b}.)
The semi-classical value of the Fermi energy,
$\lambda_{\rm sc}$, can
be determined from the particle number 
equation $N_{\rm sc}(\lambda_{\rm sc})=N$.

The semi-classical level density  
is defined as
\begin{equation}\label{gsc1}
g_{\rm sc}(\varepsilon) = \frac {dN_{\rm sc}(\varepsilon)} {d\varepsilon}.
\end{equation}
Here, it is worth reminding  that
the semi-classical level density is defined only for
$\varepsilon>V_0$, 
where $V_0$ denotes the bottom of the potential well.
That is,  $g_{\rm sc}(\varepsilon)$=0 if $\varepsilon<V_0$.
An explicit expression for $g_{\rm sc}(\varepsilon)$ in
 terms of a WK
expansion can be found in, e.g.,  Ref.~\cite{Jen74}.

It is to  be noted that Eqs.~(\ref{denmat}-\ref{gsc1}) 
are valid
for any smooth potential regardless whether it is 
infinite {\em or not}. Often, 
the semi-classical level density is 
defined in terms of the
partition function
 $Z(\beta) = \int
C(\bbox{r},\beta )\ d^3r$.
However, there is a difficulty: $Z(\beta)$
diverges for finite potentials \cite{Jen74}.
In practical calculations it is possible to 
overcome this serious  problem 
by an appropriate modification of the
potential at large disstances.
Formally,
this procedure is equivalent to placing an 
infinite potential well
(a box with soft walls)  at very large distances
from the classical region \cite{Jen74}.

Since 
the particle-number integral  (\ref{Npart1}) 
depends only on the potential in the classical region, where
 $V_0$$<$$\varepsilon$$<$$V_{\rm B}$,
the quantity $N_{\rm sc}(\varepsilon)$
is  well defined for energies
which are  not too  close to the top of the potential
well $V_{\rm B}$ where the semi-classical expansion breaks down
(see Appendix~\ref{appA}). Hence
the total number of particles can be calculated
without the explicit
use of the partition function; there is no need at all to put
the system into a box in the case of a finite potential.

The total energy of the system of non-interacting particles is
\begin{equation}\label{esc}
E=
\int_{-\infty}^{\lambda}\varepsilon g(\varepsilon)d\varepsilon=
\lambda N-\int_{-\infty}^{\lambda}
N(\varepsilon)d\varepsilon.
\end{equation}
By means of Eqs.~(\ref{totN}) and (\ref{partden}), $E$
can be written as
\begin{equation}\label{etotal}
E=\lambda N-
\int {\cal L}^{-1}_{\beta\rightarrow\lambda}
\left ( {C(\bbox{r},\beta)\over \beta^2}\right )d^3r.
\end{equation}
Here it is assumed that the order of the integration with respect to
$\varepsilon$ (and $\bbox{r}$) and the inverse Laplace transform can be
interchanged.
Using  Eq.~(\ref{etotal}),  the
semi-classical smoothed binding energy can be written as
\begin{equation}\label{etotsc}
\tilde E_{\rm sc}=\lambda_{\rm sc}N
-\int {\cal L}^{-1}_{\beta\rightarrow\lambda_{\rm sc}}
\left ( {C_{\rm sc}(\bbox{r},\beta)\over \beta^2}\right )d^3r,
\end{equation}
where the notation $C_{\rm sc}(\bbox{r},\beta)$ means that only the terms
which are not a higher
order than $\hbar$ are kept in the WK expansion of $C(\bbox{r},\beta)$.
As it was discussed in \cite{Jen75b}, for the determination
of  $\lambda_{\rm sc}$ it is enough to keep the terms of order
$\hbar^{-1}$ in Eq.(\ref{denmat}).
The explicit
expression for the smoothed energy is quite lengthy and can be found in  Ref.~\cite{Jen75b}.

\section{Generalized Shell-Correction Method}\label{newtreatment}

The impact of the particle continuum on shell corrections has been
investigated numerically in Ref.~\cite{Ro72} for neutrons
in $^{208}$Pb and $^{298}$114
by explicit  calculation
of the continuum part of the level density, Eq.~(\ref{e4}).
They  have shown that,
 by taking into account contributions 
from the neutron continuum up to $\sim$100 MeV in $^{208}$Pb, the
plateau condition (\ref{plateau}) could be  met
(see Ref.~\cite{San97} for an updated discussion 
of the continuum contribution in $^{208}$Pb).
Based on this early exercise,
it was generally {\em assumed}  that the plateau condition could
be fulfilled
for finite potentials provided that
 the continuum part was included. A systematic
study of this assumption is given in Sec.~\ref{plc}.

The lack of systematic studies using the continuum level density is
due to the fact that these
calculation are  quite cumbersome.
Except for some special cases, the solution of the radial
differential equation can be done only numerically.
The calculation of the phase shifts and the continuum level
 density along the real energy axis
should be carried out with  great care.
In general, the use of an extremely fine energy step is required
in order to collect contributions from  narrow resonances.
Recently, this difficulty has been overcome by applying
a new method which uses Gamow states \cite{San97}.
A Gamow resonance is a generalized eigenstate of the
radial Schr\"odinger
equation corresponding to a complex energy eigenvalue $w_i =\varepsilon_i-i \Gamma_i$
(for bound states $\Gamma_i$=0).
 The  wave function of a Gamow resonance is regular at $r$=0
and has purely
outgoing asymptotics with a discrete complex wave number.

In the new
method (see Ref.~\cite{San97} for details ),
the smoothed
level density $\tilde g(\varepsilon)$ can be written
with the help of the Cauchy theorem as a
 sum  over  bound and resonant states, and
an integral
along a contour
 $L$ in the complex energy plane:
\begin{equation}\label{gcsmooth}
\tilde{g}(\varepsilon) =
\sum_{i}f\left(\frac{\varepsilon-w_i}{\gamma}\right)
+\int_{L}dw~g_{\rm c}(w)
f\left(\frac{\varepsilon-w}{\gamma}\right).
\end{equation}
In Eq.~(\ref{gcsmooth}), the summation runs over all the bound states and
those resonances which are above the contour $L$ and below the real
energy axis.

Thanks to the Cauchy theorem, the level density
 as given by  Eq.~(\ref{gcsmooth})
has only real part
(the imaginary parts 
of the two terms in the
r.h.s. of Eq.~(\ref{gcsmooth}) 
should  cancel out exactly). Furtermore, $\tilde{g}(\varepsilon)$ 
should be  independent of the shape of the
contour $L$. 
However, since the calculations are carried out numerically,
the exact cancellation
of the imaginary parts is always slightly violated  and, in addition,
the sum of the real parts
slightly depends  on the shape of the contour. 
For example, if the contour
$L$ is extended to the area of the complex energy plane 
with large imaginary energies 
(Im$(w)$$<$--5 MeV) then
broad resonances,  i.e.,  those 
with large  $\Gamma_i$-values
 should be included. As a result,  both
terms in the
r.h.s. of Eq.~(\ref{gcsmooth}) would acquire
large imaginary parts
which would not
cancel completely due to the numerical errors.
Therefore, the best strategy is to choose the contour
in such a way that it would include
only  narrow resonances. With this choice the 
final result is practically independent
of the shape of the contour, and the imaginary part of $\tilde{g}(\varepsilon)$
is negligible (it is the order of $10^{-4}$ or less). Three  examples of
contours $L$ are shown in Fig.~\ref{ws}.

With  a reasonable choice for  the contour $L$, the Gamow resonances give the major
contribution from the  continuum;
the contour integral gives the remaining (small) part. From the smoothed level density (\ref{gcsmooth}) one can determine
$\tilde\lambda$ and $\tilde{E}_{\rm s.p.}$ using Eqs.~(\ref{ltilde}) and
(\ref{Esmoth}), respectively.

It is worth noting that another,  commonly used  method of
calculating the continuum level density is based on the discretization
 procedure. Here, one assumes that the nucleus is placed inside a very large
box (cf. the discussion
in Sec.~\ref{semiclassical}
on the application of the
semi-classical expansion to finite potentials). Since the properties of
the nucleus itself must not depend
on the  box size, one has to re-normalize the level density by subtracting the
contribution from the free-gas states
\cite{Shl92,Tub79,Ker81,Dea85,Shl97}. In Refs.~\cite{Shl92,Shl97}, the discretization
method was applied to investigate the accuracy of the semi-classical
expression (\ref{gsc1}) for several commonly used potentials, and a good
agreement was found in all cases.

\section{Results}\label{results}

\subsection{Details of calculations}

In the actual calculations,  we have used the average  Woods-Saxon
(WS) potential, which contains a central part, a spin-orbit term,
and a
Coulomb potential for protons.
The Coulomb potential has been  assumed to be that of a charge
$(Z-1)e$ distributed with the diffused
  charge density. We employed the set of WS
parameters introduced in Ref.~\cite{[Dud81]}, and the
charge density  form factor  was taken in the  WS form.
(See Ref.~\cite{Naz94} for details pertaining to the
single-particle model.)

The poles of the $S$-matrix, i.e. the
 eigenvalues $w_i$,
 have been  calculated by solving the radial equation
 numerically
using the computer code GAMOW
\cite{[Ve82]}. The contour $L$ has been chosen to lie
 far from these poles.
 This ensures that the
energy dependence of phase shifts
along the path is smooth; hence one can
 use relatively large energy steps.
The phase shifts of  scattering states  along the
path $L$  have been calculated by solving the radial equation numerically
using the unpublished code ZSCAT in which the complex Coulomb routine
COULCC was used\cite{[Ba]}.

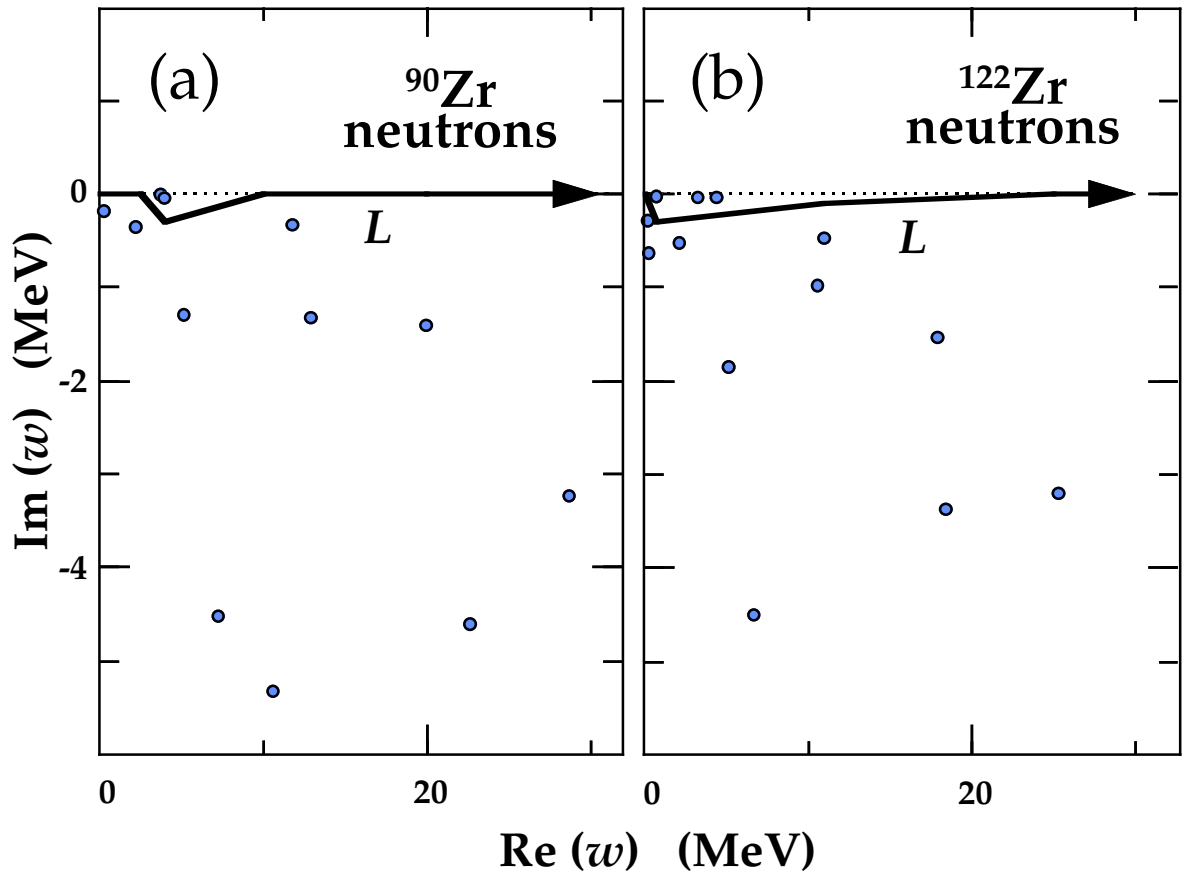
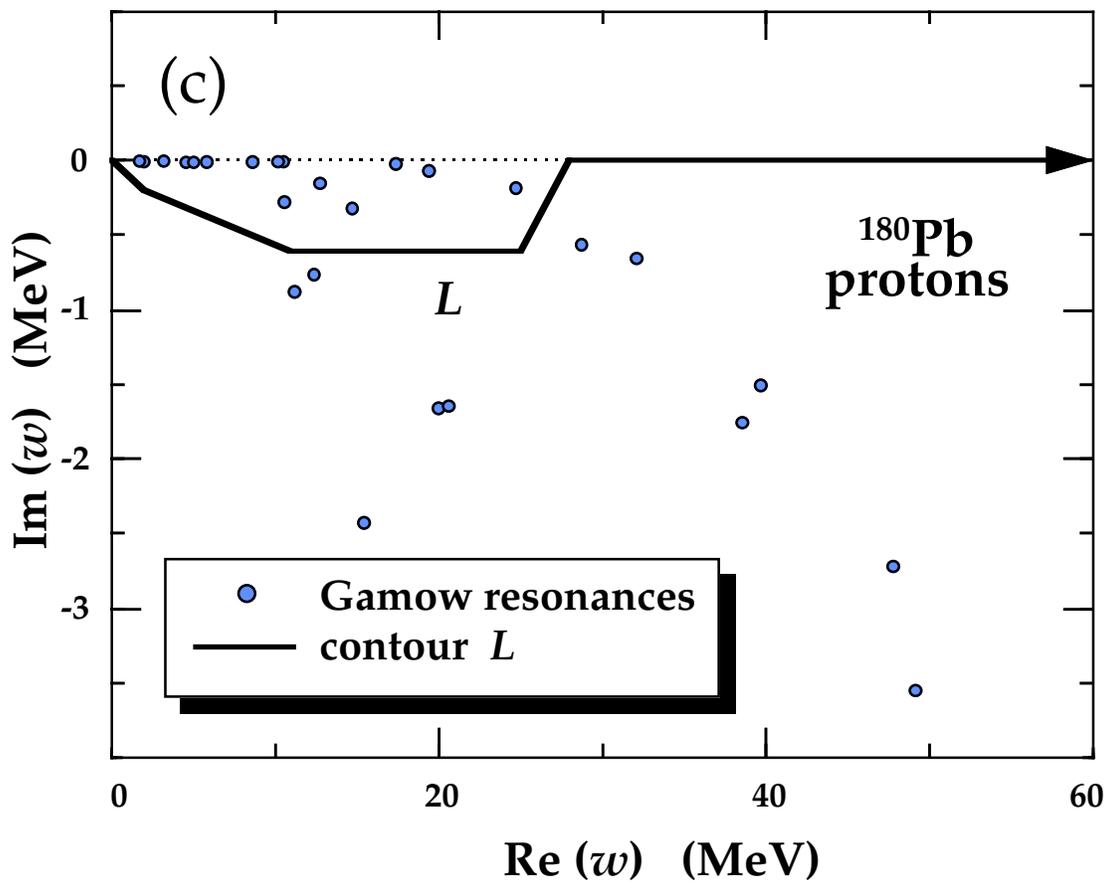
\begin{figure}[t]
\caption{The distribution of Gamow energy eigenvalues, $w_i$
in the (Re$(w)$, Im$(w)$) plane for  (a) the stable nucleus $^{90}$Zr
(neutron eigenvalues), (b)
neutron drip-line nucleus $^{122}$Zr (neutron eigenvalues), and
(c) proton-rich nucleus
$^{180}$Pb (proton eigenvalues). The contours
$L$ used  in Eq.~(\protect\ref{gcsmooth}) to calculate the
smoothed level density in the generalized shell-correction method
are also shown. Only the Gamow states with $w_i$ lying above the contour
are included in the leading term in Eq.~(\protect\ref{gcsmooth}).
}
\label{ws}
\end{figure}

As an illustrative example, the distribution of eigenvalues
 $w_i$ for the stable nucleus $^{90}$Zr
(neutrons),
neutron drip-line nucleus $^{122}$Zr (neutrons), and proton-rich nucleus
$^{180}$Pb (protons) is shown in Fig.~\ref{ws}, together with the contours
$L$ used  for calculating the level density (\ref{gcsmooth}).
In the case of $^{90}$Zr, there are
four poles close to the $\varepsilon=0$ threshold. They are:
$p_{1/2}$ ($w$=0.2$-i$0.19 MeV),
$f_{5/2}$ ($w$=2.1$-i$0.34 MeV),
$i_{13/2}$ ($w$=3.7$-i$0.004 MeV),
and
$h_{9/2}$ ($w$=3.9$-i$0.03 MeV). 
As seen in Fig.~\ref{ws}, only
the Gamow states with the smallest widths, i.e., $i_{13/2}$ and $h_{9/2}$,
have been considered
explicitly in the level density calculations; a 
 contribution from
the remaining eigenstates has been accounted
for by the integral along the path $L$, i.e., by
the second term in Eq.~(\ref{gcsmooth}). For the
neutron-rich nucleus $^{122}$Zr, the number of near-threshold Gamow states
increases. Here, three Gamow resonances:
$f_{7/2}$ ($w$=0.7$-i$0.02 MeV),
$h_{9/2}$ ($w$=3.2$-i$0.03 MeV), and
$i_{13/2}$ ($w$=4.5$-i$0.02 MeV) have been
used to define the contour that includes them in Fig. \ref{ws}b.
As seen in Fig.~\ref{ws}(c), the distribution of the proton Gamow
eigenvalues  in $^{180}$Pb is different. Due to the presence
of the Coulomb barrier, there appear many very narrow resonances
even at relatively high energies, i.e. above  10\,MeV; therefore
we have chosen in this case a contour that includes these narrow resonances.

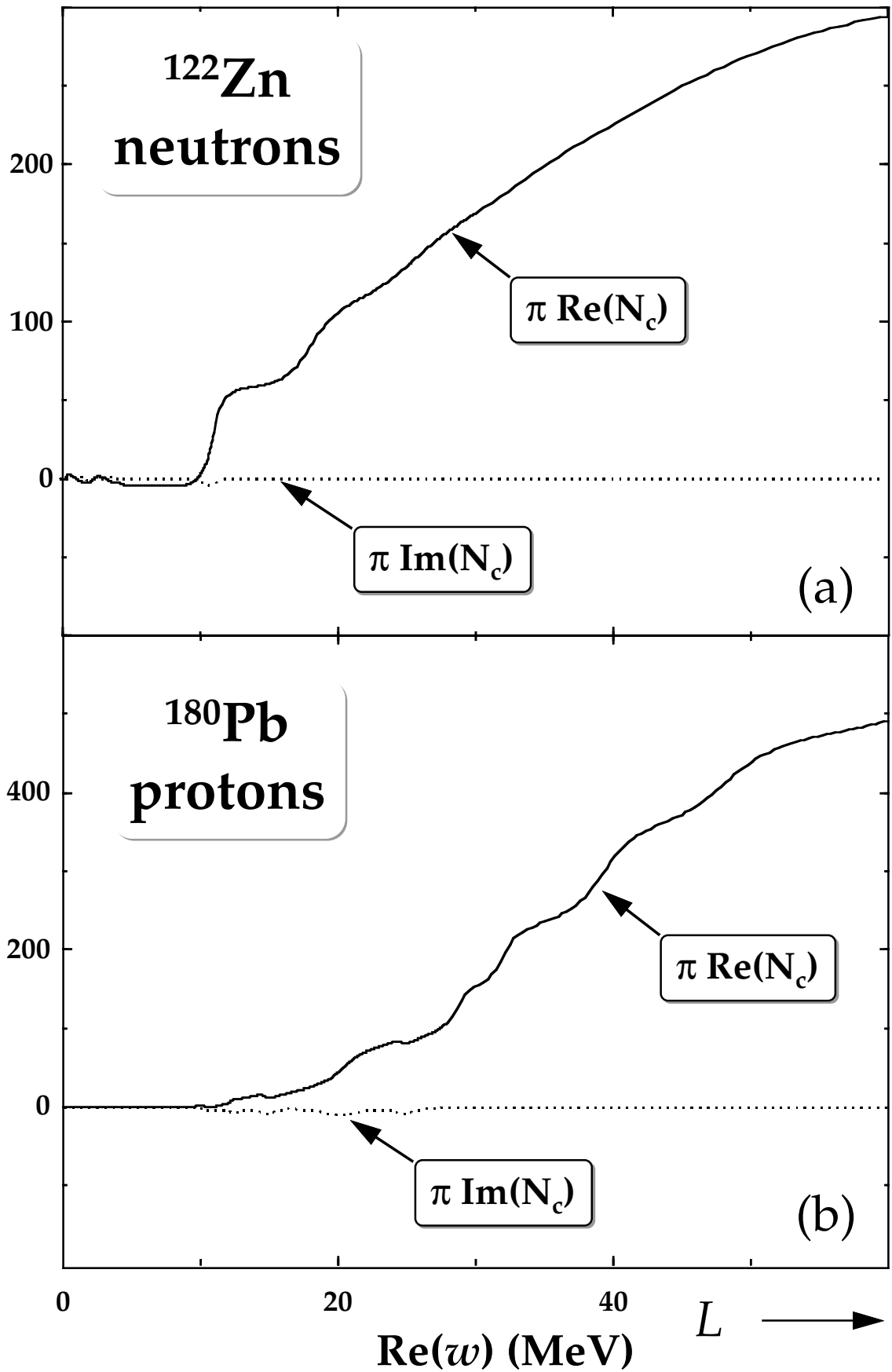
\begin{figure}[t]
\caption{The energy dependence of
 the ``continuum particle number" (\protect\ref{Nc})
for the neutrons in  $^{122}$Zr (a)
and the protons in  $^{180}$Pb (b)
along the contours $L$ of Fig.~\protect\ref{ws}(b,c).
The energy dependence of $N_{\rm c}$
along the path is very gradual. The fluctuations in $N_{\rm c}$ can
be attributed to the presence of near-lying Gamow states.
}
\label{ps}
\end{figure}

The energy dependence of neutron phase shifts
in  $^{122}$Zr and
proton phase shifts in
$^{180}$Pb along the contour $L$
 is illustrated in Fig.~\ref{ps}. Here are shown
the real and imaginary parts of
\begin{equation}\label{Nc}
N_{\rm c}(w) ={1\over \pi}
\sum_{l,j}(2j+1)\delta_{lj}(w).
\end{equation}
According to Eq.~(\ref{e4}), $N_{\rm c}$
can be interpreted as the ``continuum particle number" along
the contour. As expected, the energy dependence of $N_{\rm c}$
along the path is very smooth. It is also seen that the imaginary
part of $N_{\rm c}$ is small since the contour does not go far from
the real energy axis.

The WK calculations
in this paper follow those of Ref.~\cite{Naz94} except for
the treatment of the proton average field.
 Since the combined WS and Coulomb potentials
give rise to a non-monotonic field, in this study
we had to employ a modified
treatment  of some of the terms in the  WK expansion
according to  Ref.~\cite{Dut87}.

\subsection{Modified plateau condition}\label{plc}

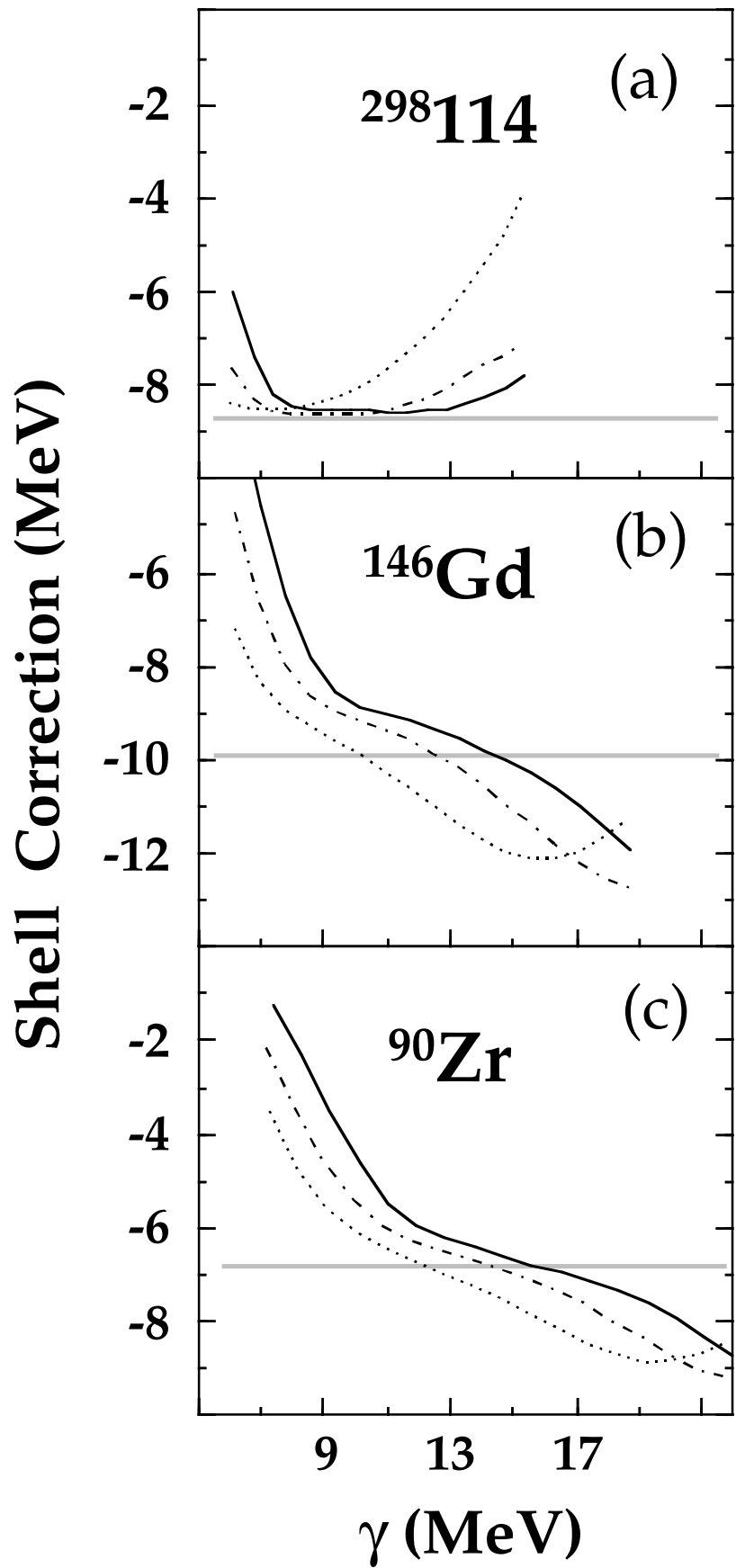
\begin{figure}[t]
\caption{Shell correction for the  neutrons
in  (a) $^{298}114$, (b)  $^{146}$Gd,  and (c) $^{90}$Zr
obtained using the generalized Strutinsky averaging procedure
as a function of the smoothing range parameter $\gamma$
for various  orders of the curvature correction:  $p$=8
(dotted line),
$p$=12 (dot-dashed line),
and $p$=16 (solid line).
The continuum contribution to the level density has been
calculated using the method described in Ref.~\protect\cite{San97}.
The gray line shows the result of the semi-classical
Wigner-Kirkwood approach.}
\label{Fig1}
\end{figure}

In order to check the dependence
of the smoothed single-particle energy on
$\gamma$ and $p$, we made systematic  calculations
of the shell energy by varying
these parameters
 within reasonable ranges.
Figure~\ref{Fig1} shows three typical examples of our
analysis of neutron shell corrections.
The  plateau condition is   satisfied fairly well
for the super-heavy nucleus $^{298}114_{184}$. (This
nucleus was previously studied in Ref.~\cite{Ro72}.) Here,
for each $p$ value,  $\delta E$ possesses a local minimum in $\gamma$,
and the minimum energy  changes little with $p$.
However, in the  cases of
 $^{146}$Gd  and  $^{90}$Zr,
it is impossible
to assign a  definite value to the neutron shell
correction. The situation for $^{146}$Gd  and  $^{90}$Zr
shown in
Fig.~\ref{Fig1}
should be considered as typical; the
plateau condition (\ref{plateau}) is seldom satisfied.
Therefore,
we conclude that the {\em proper treatment of the continuum level
density
 does not guarantee that the plateau condition  is fulfilled}.

In the cases where the plateau condition was approximately satisfied,
like in   Fig.~\ref{Fig1}(a),
we found
a strong correlation between the values of
$\gamma$ and $p$ . In particular, the behavior of
$\tilde{g}(\varepsilon)$ as a function of
$\varepsilon$ was found to be very similar
for different values of $p$ and $\gamma$
 corresponding to local minima in $\delta E$.
 Moreover,
the dependence of  $\tilde{g}(\varepsilon)$ was found to be
almost linear  in a wide range of $\varepsilon$ below
$\tilde{\lambda}$. We also checked that
 for the cases when the plateau condition could not be satisfied
[see, e.g.,  Figs.~\ref{Fig1}(b,c)], the
approximate linearity of
$\tilde{g}(\varepsilon)$ was valid.
It is worthwhile to point out that for the  harmonic oscillator potential,
the average level density
behaves as $\varepsilon^2$, while for the finite square well potential,
the leading terms behave as $\sqrt{\varepsilon}$ \cite{Bha71,Shl92}.
 Hence, a local linear
behavior of $\tilde{g}(\varepsilon)$ for a finite WS potential is not
 unexpected.
This observation suggests that an
alternative recipe
for defining  shell correction for finite potentials,
not based on the
plateau condition but rather on the behavior of the smooth level density,
may be possible.

It is well known that the realistic value of the smoothing parameter
has to lie in a certain energy interval \cite{Jen74}.
The value of $\gamma$  should be large enough
to wipe out shell effects in the energy  range of
a typical distance between shells:
\begin{equation}
\gamma>\hbar\Omega.
\end{equation}
On the other hand,
its upper limit is defined by the number of states considered in
the calculations, i.e.,
\begin{equation}\label{cond1}
\gamma\ll \varepsilon_{\rm max}-\tilde{\lambda},
\end{equation}
and by
 the energy distance between the Fermi level and the bottom
of the well \cite{Bra73}, i.e.,
\begin{equation}\label{cond2}
\gamma\ll \tilde{\lambda}-V_0.
\end{equation}
In practice,  the optimal value of $\gamma$ for a given $p$ is found by
the following procedure. First we choose the energy interval
  $[\varepsilon_l,\varepsilon_u]$
which is lying below $\tilde{\lambda}$ and is
wider than
the average shell distance, e.g.,
$\varepsilon_u$--$\varepsilon_l$$>$1.5\,$\hbar\Omega$.
In this energy interval, we perform the least squares fit
to the smoothed level density
assuming a linear dependence of $\tilde{g}(\varepsilon)$ on $\varepsilon$.
The search for  optimal $\gamma$
begins at a small $\gamma$  value below $\hbar\Omega$
where   the shell fluctuations are still present, and then
 $\gamma$ is gradually increased until the first minimum
in $\delta E$ is found at $\gamma_{p}$.
This  $\gamma_{p}$ corresponds to  the smallest  value
of $\gamma$ for a  given $p$
that  smooths out the shell fluctuations.
The corresponding shell correction, $\delta E=\delta
E_p$, is taken
as the optimal shell correction  for this $p$.
This procedure is  repeated for higher values of $p$.
If  variations of  $\delta E_{p}$ with
 $p$ are small,  then the mean value of  $\delta E_{p}$ represents
the shell correction obtained in this
modified  Strutinsky method. The uncertainty of the procedure
is given by the r.m.s. error   $\sigma$$\equiv$$\sigma(\delta E)$.

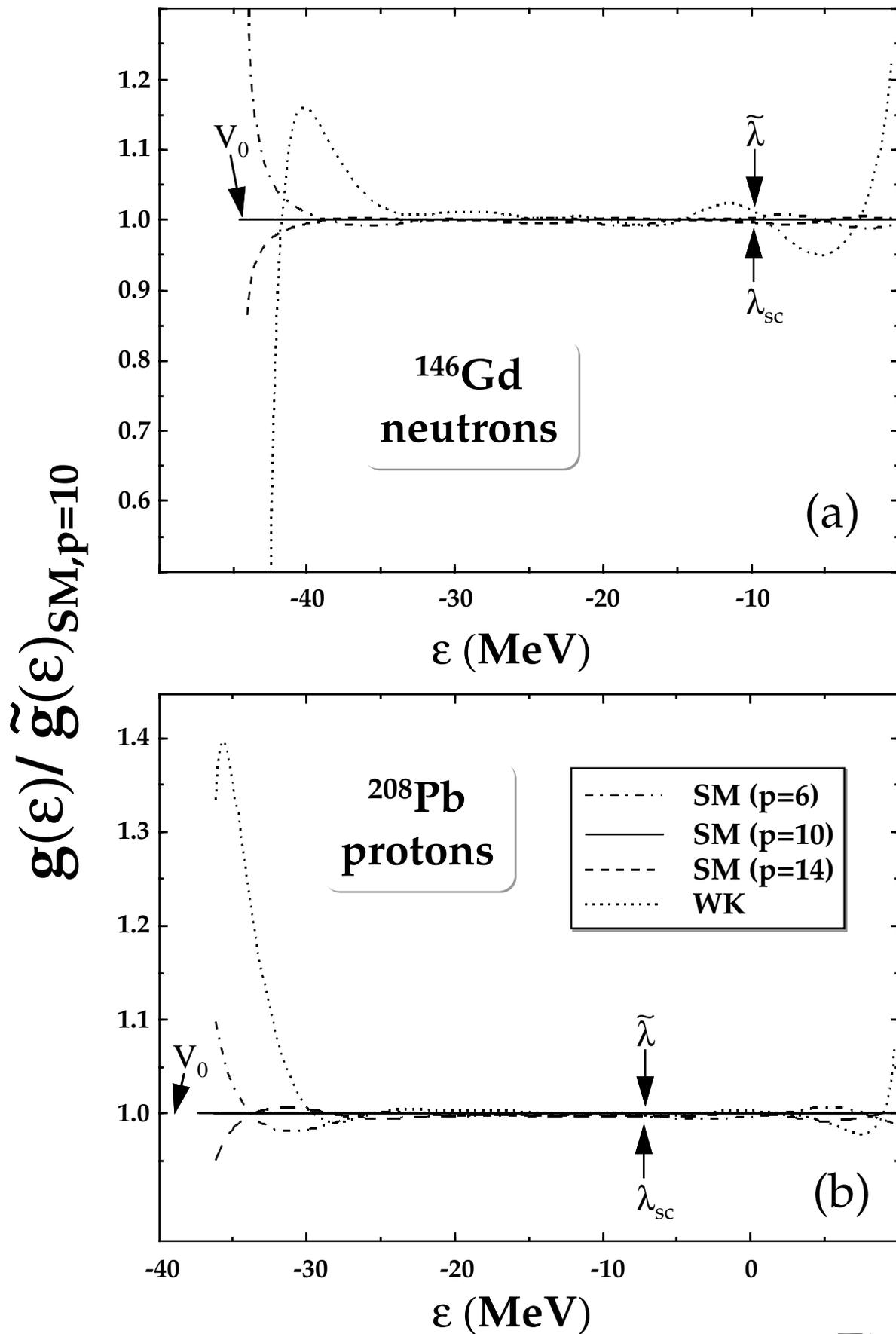
\begin{figure}[t]
\caption{Comparison of the smoothed level densities calculated using
the generalized
Strutinsky method (SM) and the
Wigner-Kirkwood method (WK)
for (a) the neutrons
in $^{146}$Gd  and  (b) the protons in $^{208}$Pb.
The  densities are normalized to the Strutinsky
density $\tilde{g}(\varepsilon)$ calculated with the
curvature correction $p$=10.
Semi-classical level densities and Strutinsky level densities
calculated with  $p$=6  and 14 are shown by dotted,
dot-dashed,  and dashed lines, respectively. It is seen that
 the result
of the Strutinsky smoothing
is practically  $p$-independent.
The Fermi levels $\tilde\lambda$ (SM) and $\lambda_{\rm sc}$ (WK)
are indicated, together with the value of the potential depth $V_0$.
}
\label{Fig2}
\end{figure}

The smoothed
level densities $\tilde{g}(\varepsilon)$, calculated using the above procedure,
are  displayed in Fig.~\ref{Fig2} for  three
different values  of $p$=6, 10, and 14.
Since $\tilde{g}(\varepsilon)$ is practically
independent of the value of $p$ we amplified the differences by presenting
in Fig.~\ref{Fig2} the ratios of the level densities to $\tilde{g}(\varepsilon)$
belonging to $p=10$.

 The shell energy is also practically
independent of the value of $p$.
Table~\ref{tab1} displays the calculated proton and neutron shell-correction
energies  and
the corresponding r.m.s. errors.
Since $\sigma$  increases when  going to lighter nuclei
where the condition (\ref{cond2}) does not hold,
we limited calculations to nuclei heavier than
 $^{40}$Ca. The calculations were performed for nuclei close
to the stability valley and for nuclei with extreme $N/Z$ ratios (both
neutron-rich and proton-rich).
It is seen that  the
r.m.s. error in
$\delta E$  is always less than
 250 keV (also for the cases such as $^{90}$Zr or $^{146}$Gd where
the plateau condition could not be met).
Another source of theoretical uncertainty lies in  the choice
of the fitting range $[\varepsilon_l,\varepsilon_u]$. In particular,
the selection  of $\varepsilon_u$ plays a role
for weakly bound nuclei with very small values of $\tilde\lambda$.
In practice, in order to guarantee that the fitting region
does not overlap with the threshold area, we have adopted the value of
$\varepsilon_u$=$\tilde\lambda-\hbar\Omega$ and
$\varepsilon_l$=$\varepsilon_u-1.5\hbar\Omega$.
In order to estimate the associated theoretical error, we have
performed a set of calculations for well-bound nuclei assuming
a larger value of
$\varepsilon_u$, namely $\varepsilon_u$=$\tilde\lambda$. The  average deviation in $\delta E$
between the two sets  of calculations is around 400\,keV. We believe
that this number represents
a fair estimate of the uncertainty of our method.

\subsection{Comparison with the semiclassical method}\label{SMWK}

As a next step, we performed the detailed comparison
of the generalized Strutinsky method with the WK expansion.
Table~\ref{tab1}  displays  the
shell correction $\delta E_{\rm sc}= E_{\rm s.p.}- \tilde{E}_{\rm sc}$
calculated using the
WK method, together with the
difference $\Delta$$\equiv$$\delta E-\delta E_{\rm sc}$.
In most cases  $\Delta$$>$0. That is,
the semi-classical method yields the average
single-particle energy  $\tilde{E}_{\rm sc}$ which is greater than
 $\tilde{E}_{\rm s.p.}$ obtained in the generalized Strutinsky method.
The average value of $\Delta$
is about 0.45 MeV and the maximal deviation is about 1.8 MeV for neutrons
and 0.9 MeV for protons.
These deviations exhibit large fluctuations with particle number,
and they
 are significantly larger than
 the uncertainty of the generalized Strutinsky smoothing procedure.
 Considering the excellent agreement
between the shell energies  calculated in  the  Strutinsky  and
the WK methods
obtained  previously \cite{Jen75b}, and the existing proof
of the equivalence between these two methods \cite{Jen73},
the large magnitudes  of $\Delta$  in Table~\ref{tab1}
 seem to be  surprising.

In order to understand this discrepancy, we analyze the behavior of
smoothed single-particle level densities obtained in both
 methods. The semi-classical level density has been obtained
by means of Eq.~(\ref{gsc1}), i.e., by calculating the derivative
of $N_{\rm sc}(\varepsilon)$.

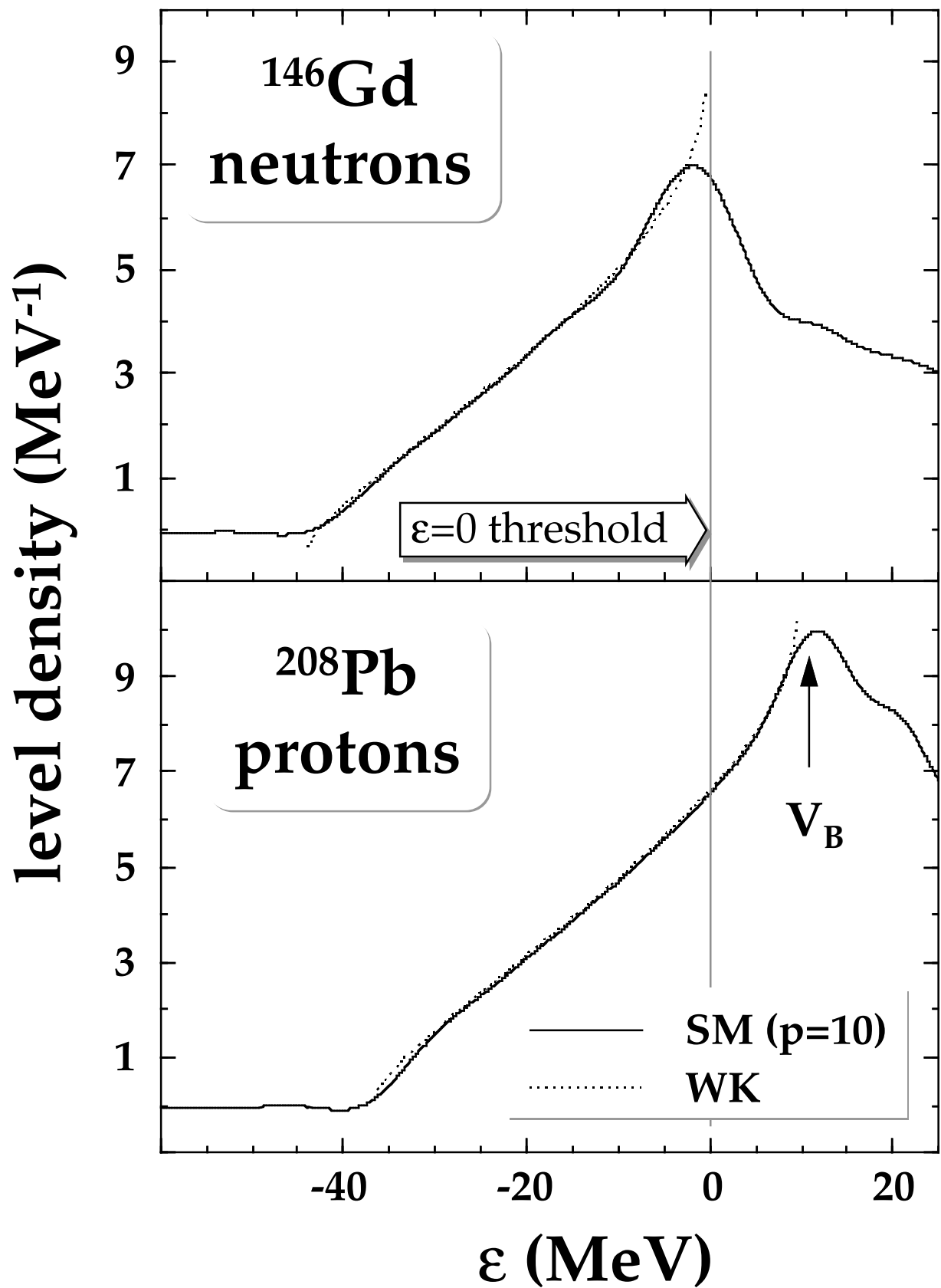
\begin{figure}[t]
\caption{Comparison of the smoothed level densities calculated using
the generalized
Strutinsky method (solid line, $p$=10 variant) and the
Wigner-Kirkwood method (dotted line) for the neutrons
in $^{146}$Gd (top) and  the protons in $^{208}$Pb (bottom).
}
\label{Fig3}
\end{figure}

Smoothed level densities $\tilde{g}(\varepsilon)$
and $g_{\rm sc}(\varepsilon)$ are compared in Figs.~\ref{Fig2} and \ref{Fig3}
for  $^{146}$Gd (neutrons) and $^{208}$Pb (protons), respectively.
The behavior of average single-particle densities
displayed in Fig.~\ref{Fig3} shows a generic
pattern characteristic of a WS-like potential well
\cite{Shl92,Shl97}.
Namely,  $\tilde{g}(\varepsilon)$ increases monotonically with $\varepsilon$
reaching the maximum value
around the $\varepsilon$=0 threshold for the neutrons (and around $V_B$, i.e.
the top of the Coulomb barrier for the protons), and then it smoothly falls
down reflecting the increasing contribution from the free-gas states.
As discussed in Sec.~\ref{plc},
there exists a wide energy region in which
$g_{\rm sc}(\varepsilon)$ increases fairly linearly with $\varepsilon$.
In general, $\tilde{g}(\varepsilon)$
and $g_{\rm sc}(\varepsilon)$ are  very close
  except for  the bottom of the potential well ($\varepsilon$$\approx$$V_0$)
and close to the top of the potential barrier.

Considering the low-energy region, there are problems with both
methods. The Strutinsky smoothed density is non-zero
for $\varepsilon<V_0$, i.e., in the classically forbidden region. Here,
the inequality (\ref{cond2}) cannot be met
and the averaging method breaks down \cite{Bra73}.
Also, there are  serious questions
regarding the applicability of the semi-classical treatment  close
to the $\varepsilon$=$V_0$ limit.
 The $\hbar^{-1}$ term in the
WK   expansion of the semi-classical level density   [related to the
second term in the integral (\ref{Npart1})]
gives rise to the singularity
around the bottom of the well. (For a square well potential
 this singularity behaves as $(\varepsilon-V_0)^{-1/2}$.)
As discussed in Ref.~\cite{Jen74},
in the strict treatment of $g_{\rm sc}(\varepsilon)$,
there appears  a correction to the level density, proportional to
the Dirac delta $\delta(\varepsilon-V_0)$. This
additional term, usually ignored in calculations,
   partly corrects for a pathological
behavior around the bottom of the well by
introducing  a small shift in the Fermi
level $\lambda_{\rm sc}$ \cite{Bha71}.
Table~\ref{tab1}  displays  the Fermi energies $\tilde{\lambda}$
and $\lambda_{\rm sc}$. Usually, the
Fermi levels are very close; for well-bound nuclei the
difference is less than 100\,keV.
Although small,
this shift partly
contributes to the calculated values of $\Delta$ (e.g.,
for protons in $^{208}$Pb).

Another  major deviation between  $\tilde{g}(\varepsilon)$
and $g_{\rm sc}(\varepsilon)$ is
seen in the energy region close to the $\varepsilon$=0 threshold
in the neutrons,  and around the top of
the proton Coulomb barrier.
Here, the reason for this abnormal  behavior is
 our semi-classical approximation.
As discussed in Appendix~\ref{appA}, the WK neutron level density
$g_{\rm sc}(\varepsilon)$ diverges as
$\ln(-\varepsilon)/\sqrt{-\varepsilon}$ when $\varepsilon\rightarrow 0$.
 For protons, the singularity is even more severe:
$g_{\rm sc}(\varepsilon)$  diverges as $(V_{\rm B}-\varepsilon)^{-1}$
around the top of the barrier.

The pathological behavior of $g_{\rm sc}(\varepsilon)$ around zero energy
is the reason  for the largest
deviations $\Delta$
seen in Table~\ref{tab1}
for the neutron case
(e.g., $\Delta$=1.39\,MeV for $^{78}$Ni, and
it is greater than 1\,MeV for the Zr isotopes
with $A$=110,122,124).
These nuclei are weakly bound
(as shown by their low  Fermi energies), and
the level density around the Fermi level, $g_{\rm sc}(\tilde{\lambda})$,
is affected by the threshold effect.

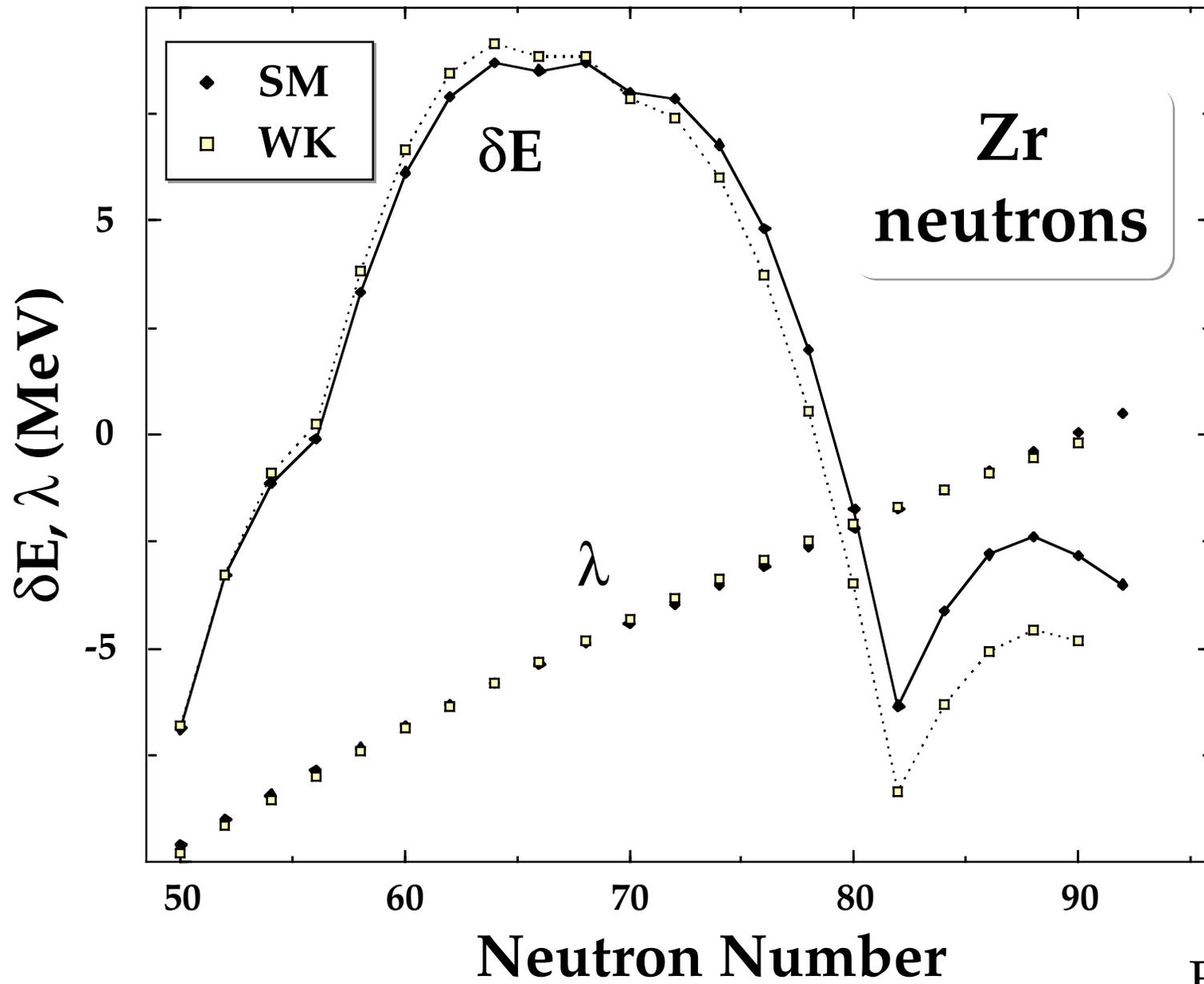
\begin{figure}[t]
\caption{Neutron shell corrections and Fermi energies
as a function of $N$ calculated in the SM and WK models
using the single-particle potential of $^{90}$Zr.
The corresponding smoothed level densities
are shown in Fig.~\protect\ref{Zr90}.
}
\label{Zrlambda}
\end{figure}

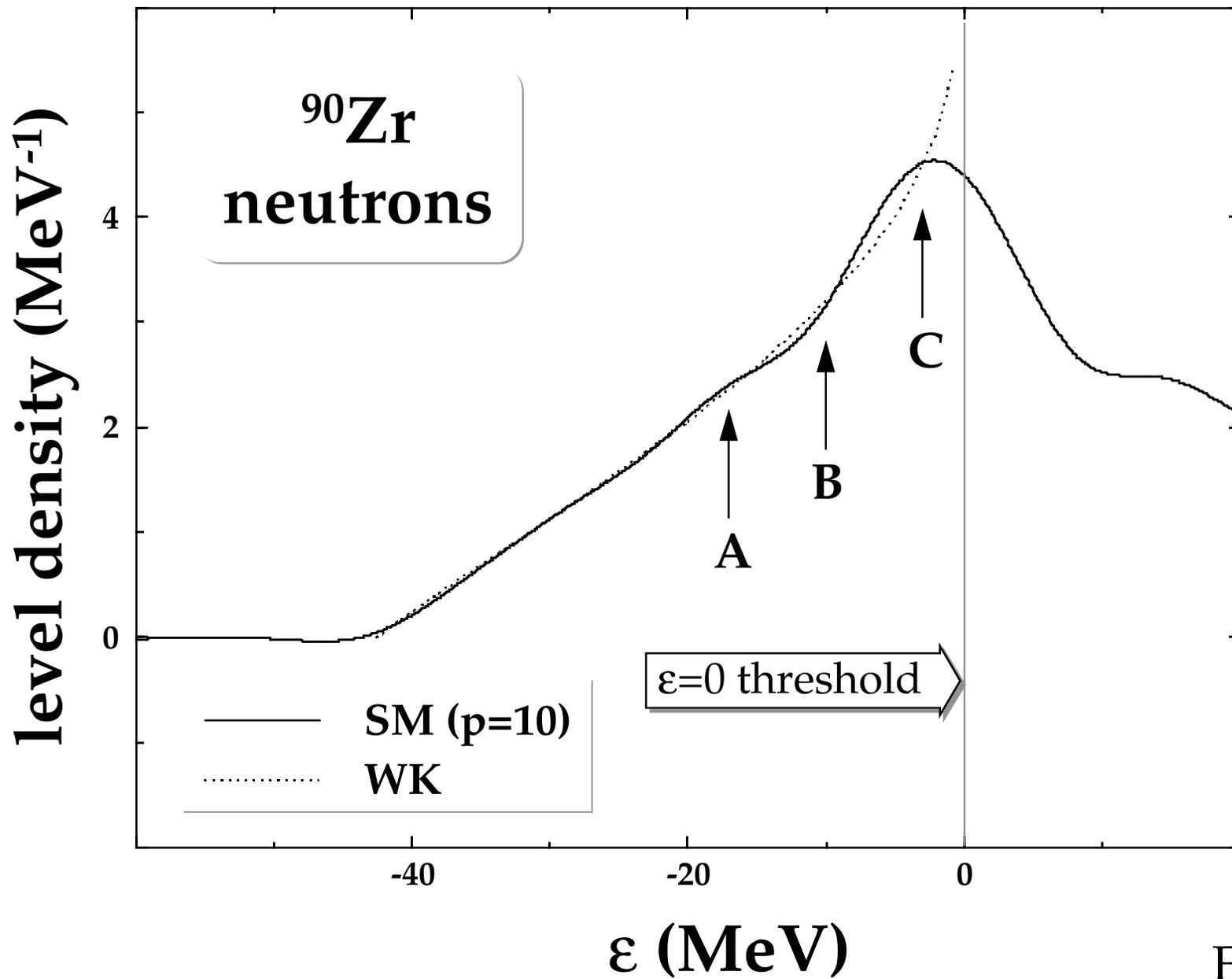
\begin{figure}[t]
\caption{Same as in Fig.~\protect\ref{Fig3} except
for the neutrons
in $^{90}$Zr. The points
at  which the difference between SM and WK level densities,
$\delta g$,
changes sign are marked by
``A", ``B", and ``C". The oscillatory behavior
of $\delta g$
is responsible
for the oscillatory behavior of $\Delta$ as a function of particle number,
as shown in Fig.~\protect\ref{Zrlambda}.
}
\label{Zr90}
\end{figure}

In order to understand the systematic behavior of $\Delta$
in Table~\ref{tab1},
neutron shell corrections and Fermi energies  for the Zr isotopes
are shown in Fig.~\protect\ref{Zrlambda}
as  functions of $N$.
The calculations were performed
using the single-particle potential
corresponding to  $^{90}$Zr.
The associated  smoothed level densities
are shown in Fig.~\ref{Zr90}. It is seen that
although the general pattern of $\delta E$ is similar
in the generalized shell-correction method and WK approach,
$\Delta$ exhibits
the oscillatory behavior as a function of particle number.
The agreement between $\delta E$ and $\delta E_{\rm sc}$
is very good up to $N$$\sim$70 ($\tilde\lambda$$\sim$--4\,MeV),
but it is spoiled at  large neutron numbers
where $\Delta$ systematically
increases. Indeed, above  point ``C"  in Fig.~\ref{Zr90},
the semi-classical level density diverges, and it does not
yield a good estimate of the shell correction.

For the protons,
the ``dangerous" threshold region
of  $g_{\rm sc}$ is shifted to higher energies
due to the presence of the Coulomb barrier (see Fig.~\ref{Fig3}).
That explains why
the  differences between the Strutinsky and WK
results are smaller for the protons than for the neutrons.
For instance,
for the proton drip-line nucleus $^{100}$Sn
the agreement between two methods is
surprisingly good in spite of the fact that
$\tilde{\lambda}$=0.72\,MeV.
Since in actual nuclei (both
particle-bound and in proton emitters) the
proton Fermi level is  significantly lower than $V_{\rm B}$,
the divergent behavior of proton $g_{\rm sc}$
around the top of the Coulomb barrier
has no practical importance.

\section{Conclusions}\label{conclusions}

This paper introduces a new method of calculating
the nuclear shell energy.  The generalized Strutinsky procedure
fully takes into account the effect of the particle continuum.
Although the traditional plateau condition can seldom be met
for finite potentials,
the proposed method makes it possible to define  shell correction
unambiguously. A conservative estimate of the uncertainty in
$\delta E$ obtained
using the new smoothing procedure
is $\sim$400\,keV. This error can be considered as small.

In most cases, the results
of the generalized Strutinsky procedure are in  good  agreement with those
of the semi-classical WK method.
Significant deviations have been obtained for neutron-rich nuclei
for which
the neutron Fermi energy is low
($\lambda$$>$--4 MeV). This discrepancy has been
tracked back  to the
singularity in  the WK level density around the top of the
potential barrier.
The density $\tilde{g}(\varepsilon)$ obtained in the
generalized Strutinsky method
nicely interpolates through the threshold  region
(see also Refs.~\cite{Shl92,Shl97}).
Other advantages of the new method are:
 (i)  its  applicability
to  potentials
with discontinuous derivatives (e.g., the Coulomb potential of a
uniform  charge distribution  and  a folded-Yukawa potential) where
the standard WK expansion cannot be carried out, and (ii)
simple generalization to the deformed case where the semiclassical
expansion becomes  awkward \cite{Naz94}.

Finally, let us comment on the difference  $\Delta$
between shell corrections obtained in both methods.
There are  several factors which contribute to $\Delta$.
In addition to the threshold anomaly  mentioned above, other
factors are:
(i) the shift between the Fermi levels caused by the
different behavior of the level densities at the bottom of the
potential well,
(ii) the assumption of the local linearity of the smoothed level
density,  and  (iii) the systematic errors
accumulated during  numerical calculations.
For protons,  our calculations give
$|\Delta|$$<$900\,keV  in all cases considered. Here, the main source
of the  difference is the deviation between the level
densities around the bottom of the potential well, i.e., factor (i).
For neutrons with
$\tilde{\lambda}$$<$--4 MeV, the value of
$|\Delta|$ is even smaller: $|\Delta|$$<$600\,keV. The largest deviations approaching 2\,MeV
have been obtained for the neutron drip-line nuclei such as $^{122}$Zr.
Considering the  analysis presented in this study, it has to be concluded that the excellent  agreement
found in  Ref.~\cite{Jen75b}
($|\Delta|$$\sim$100\,keV) is fortuitous. According to
our  results  in Table~\ref{tab1}, for nuclei discussed
by these authors, i.e., $^{208}$Pb and $^{208}$114,
the difference $\Delta$ is indeed very  small. However,
in other cases deviations are larger.

\acknowledgments

This research was supported in part by
 the Hungarian National Research Fund
(OTKA T17298), the Swedish Royal Academy of Sciences,
 the U.S. Department of Energy
under Contract Nos. DE-FG02-96ER40963 (University of Tennessee),
DE-FG05-87ER40361 (Joint Institute for Heavy Ion Research),
DE-AC05-96OR22464 with Lockheed Martin Energy Research Corp. (Oak
Ridge National Laboratory), the Institute of Atomic Physics, Bucharest,
and  by the Polish Committee for
Scientific Research.

\appendix
\section{Near-threshold behavior of the semiclassical level density}
\label{appA}

In the WK method, the semiclassical level density $g_{\rm sc}(\varepsilon)$
can be written as
\begin{equation}\label{ggg}
g_{\rm sc}(\varepsilon) = g_{\rm TF}(\varepsilon) + g_{-1}(\varepsilon),
\end{equation}
where $g_{\rm TF}(\varepsilon)$ is the Thomas-Fermi (TF) level density
and $g_{-1}(\varepsilon)$ is the  WK correction term
(of the order of  $\hbar^{-1}$).
By employing Eqs.~(\ref{gsc1}) and (\ref{Npart1}), one can
write  $g_{\rm sc}(\varepsilon)$ as a derivative
of the particle number with respect to $\varepsilon$:
\begin{equation}\label{gg}
g_{\rm TF}(\varepsilon)=\frac{dN_{\rm TF}}{d\varepsilon},
~~g_{-1}(\varepsilon)=\frac{dN_{-1}}{d\varepsilon}.
\end{equation}
In Eq.~(\ref{gg}), $N_{\rm TF}$ is the TF particle number,
\begin{equation}\label{NTF}
N_{\rm TF}(\varepsilon) = \frac{4}{3\pi}\left(\frac{2M}{\hbar^2}\right)^{3\over 2}
\int^{r_{\rm sc}(\varepsilon)}  (\varepsilon-V)^{3\over 2}  r^2 dr,
\end{equation}
while the $\hbar^{-1}$ WK term  is
\begin{equation}\label{Ncorr}
N_{-1}(\varepsilon) = -\frac{1}{12\pi}\left(\frac{2M}{\hbar^2}\right)^{1\over 2}
\int^{r_{\rm sc}(\varepsilon)}  \frac{\nabla^2 V}{\sqrt{\varepsilon-V}} r^2 dr.
\end{equation}
As usual, the classical turning point is
 defined by the relation
$V(r_{\rm sc}(\varepsilon))=\varepsilon$.

The pathological behavior of $g_{\rm sc}(\varepsilon)$ close to the top of the
potential barrier can be attributed to the  singularity  in
the $g_{-1}(\varepsilon)$ term. To examine this divergence, let us consider
the integral
\begin{equation}
I(\varepsilon) \equiv \int^{r_{\rm sc}(\varepsilon)}_{\tilde{r}}
 \frac{\nabla^2 V}{\sqrt{\varepsilon-V}}
 r^2 dr,
\end{equation}
where it has been  assumed that $V(r)$  is an increasing  function of $r$
on an  interval  $[\tilde{r}, r_{\rm sc}]$, and $\tilde{r}$ is a fixed
radius  ($\tilde{r}< r_{\rm sc}$).

By substituting $x=V(r)$, $I(\varepsilon)$ can be written as:
\begin{equation}\label{II}
I(\varepsilon) = \int^{\varepsilon}_{\tilde{x}}
\frac{\eta(x)}{\sqrt{\varepsilon-x}}
 dx,
\end{equation}
where $\tilde{x}=V(\tilde{r})$ and
\begin{equation}\label{eta}
\eta(x) \equiv r^2\frac{\nabla^2 V}{V'}|_x =
\left(r^2\frac{V''}{V'}+2r\right)|_x.
\end{equation}
The singularity at $x=\varepsilon$
in the integrand in Eq.~(\ref{II}) can be eliminated  by performing a partial integration:
\begin{equation}\label{II1}
I(\varepsilon) =  2\eta(\tilde{x})\sqrt{\varepsilon-\tilde{x}}  +2
\int^{\varepsilon}_{\tilde{x}} \eta'(x)\sqrt{\varepsilon-x} dx.
\end{equation}
The first term in Eq.~(\ref{II1}) does not cause any problems around the
particle threshold and can be neglected in the following.
Hence
the behavior of $g_{-1}(\varepsilon)$ around the top of the barrier is
governed by the integral
\begin{equation}\label{Ifinal}
I'(\varepsilon) \approx  \int^{\varepsilon}_{\tilde{x}} \frac{\eta'(x)}{\sqrt{\varepsilon-x}}dx.
\end{equation}

\subsection{Woods Saxon potential: neutron case}

For neutrons in a WS potential, the particle threshold appears at $\varepsilon$=0.
In the vicinity of the threshold, the potential energy can be approximated by
\begin{equation}
V(r)\approx \tilde{x}\exp\left({-\frac{r-\tilde{r}}{a}}\right),
\end{equation}
and the classical radius  is
\begin{equation}
r(x)=\tilde{r}+a\ln\left(\frac{\tilde{x}}{x}\right).
\end{equation}

In the limit $\varepsilon \rightarrow 0$,
 the function $\eta(x)$ in Eq.~(\ref{eta}) can be written as:
\begin{equation}
\eta(x)=2r(x)-r^2(x)/a \approx -r^2(x)/a.
\end{equation}
Hence
\begin{equation}
\eta'(x) \approx  2\frac{r(x)}{x}
\end{equation}
and
\begin{equation}\label{IWS}
I'(\varepsilon) \approx 2 \int^{\varepsilon}_{\tilde{x}}
\frac{\ln\left(\frac{\tilde{x}}{x}\right)}{x\sqrt{\varepsilon-x}}dx.
\end{equation}
The above integral can be easily calculated. Around
$\varepsilon$=0, it behaves as
\begin{equation}\label{g_n}
\ln\left(\frac{\varepsilon}{\tilde{x}}\right)/
{\sqrt{
\frac{\varepsilon}{\tilde{x}}
}},
\end{equation}
and this is the
asymptotic behavior of $g_{\rm sc}(\varepsilon)$ around the
neutron  threshold.

The behavior
of $g_{\rm sc}(\varepsilon)$ for the neutrons in $^{120}$Sn
at $\varepsilon$$\sim$0
is displayed  in Fig.~\ref{ge}(a). It
is seen that
the semiclassical density diverges according
to the law given by Eq.~(\ref{g_n}).

\subsection{Finite potential barrier: proton case}

For  potentials with finite barriers,
such as the sum of WS and Coulomb potentials,
the particle threshold appears at the top
of the barrier, $\varepsilon$=$V_{\rm B}$.
Around the barrier top, the potential energy can be
expanded as
\begin{equation}
V(r)\approx V_{\rm B} - \alpha \frac{1}{2} (r-r_{\rm B})^2,
\end{equation}
where $\alpha=-V''(r_{\rm B})$.

For $\varepsilon \approx V_{\rm B}$,
the function $\eta(x)$ in Eq.~(\ref{eta}) can be written as:
\begin{equation}
\eta(x)  \approx -\frac{r^2(x)}{r_{\rm B}-r(x)}.
\end{equation}
Consequently,
\begin{equation}
\eta'(x) \approx  -\frac{r^2(x)}{\alpha [r_{\rm B}-r(x)]^3},
\end{equation}
and the leading term in $I'(\varepsilon)$ takes the form
\begin{equation}\label{IWS1}
I'(\varepsilon) \rightarrow  \int^{\varepsilon}_{\tilde{x}} \frac{1}
{(V_{\rm B}-x)^{3/2}\sqrt{\varepsilon-x}}dx.
\end{equation}
The above integral can be easily evaluated. Around
$\varepsilon$=$V_{\rm B}$ it behaves as
\begin{equation}\label{g_p}
\frac{1}{V_{\rm B}-\varepsilon},
\end{equation}
and this gives the
asymptotic behavior of $g_{\rm sc}(\varepsilon)$ around the
top of the Coulomb barrier.

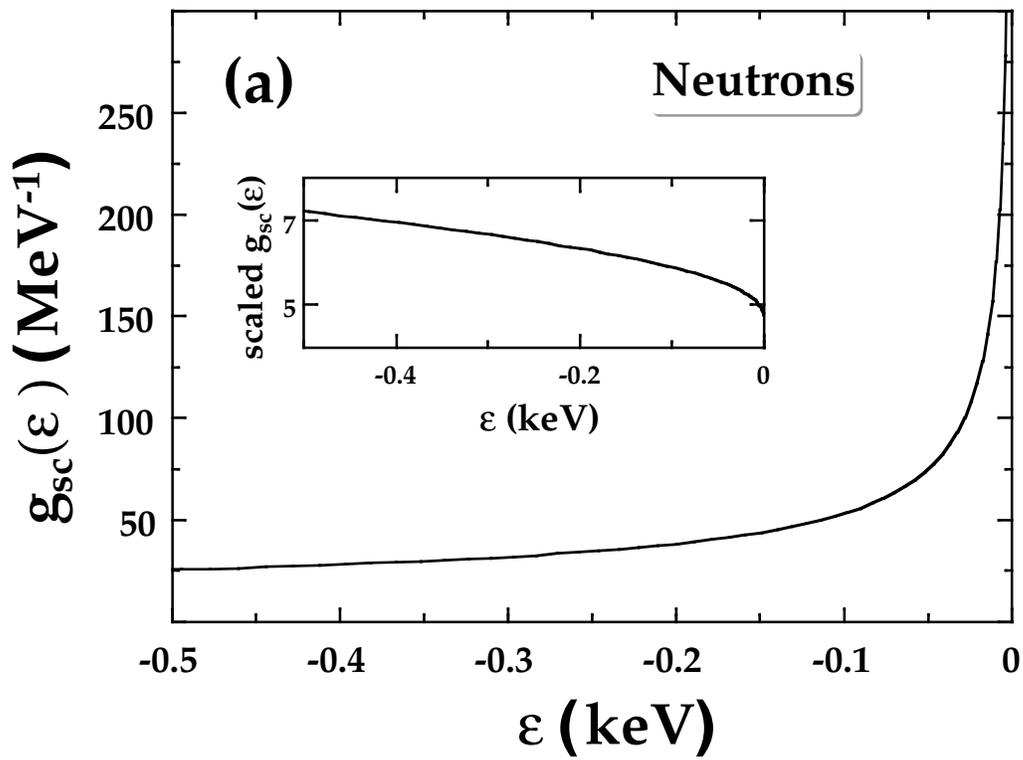
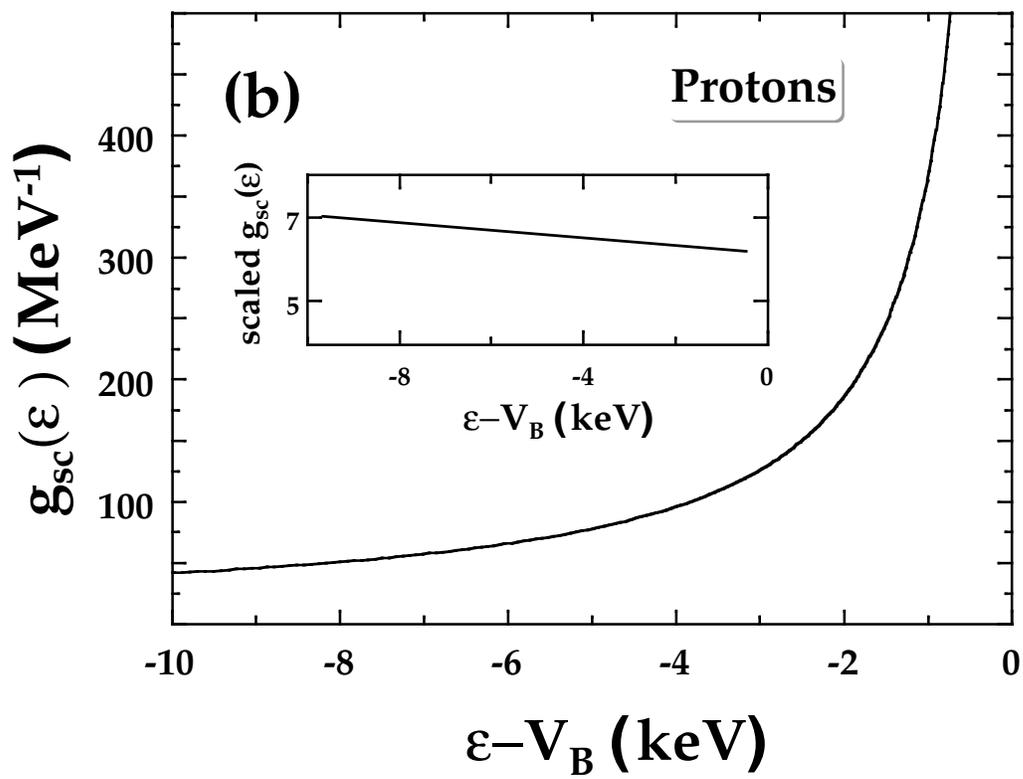
\begin{figure}[t]
\caption{The divergent behavior
of $g_{\rm sc}(\varepsilon)$ for the neutrons
(a) and the protons (b)
in $^{120}$Sn around the
particle threshold.
It is seen that
 the semi-classical approximation
breaks down in the vicinity of the threshold.
The densities scaled according to
Eqs.~(\protect\ref{g_n}):
$1000 g_{\rm sc}(\varepsilon)/[\ln(\tilde{\varepsilon})/
\protect\sqrt{\tilde{\varepsilon}}]$
(where $\tilde{\varepsilon} =\varepsilon/V_0$), and
  Eq.~(\protect\ref{g_p}):
$1000 g_{\rm sc}(\varepsilon)/[1/\tilde{\varepsilon}]$
(where $\tilde{\varepsilon}=(\varepsilon-V_{\rm B})/V_0$)
are shown in the insets.
}
\label{ge}
\end{figure}

The behavior
of $g_{\rm sc}(\varepsilon)$ for the  protons in $^{120}$Sn around
$\varepsilon$=$V_{\rm B}$
is displayed  in Fig.~\ref{ge}(b). It
is seen that  around the top of the barrier
the semi-classical density diverges according
to the law given by Eq.~(\ref{g_p}). Of course, the WK
contribution  (\ref{Ncorr})
to the particle number also diverges when $\varepsilon \rightarrow V_{\rm B}$.
This result is by no means surprising; the semi-classical approximation
breaks down if the gradient of the potential at the turning point vanishes,
and this happens
precisely around the top of the barrier.

\newpage

\begin{table}
\caption{Shell correction $\delta E$ , the  r.m.s.
error
$\sigma$, and the Fermi level $\tilde{\lambda}$ calculated
using the generalized Strutinsky method
with continuum. The corresponding semi-classical quantities:
shell correction $\delta E_{\rm sc}$ and Fermi level $\lambda_{\rm sc}$
 are also shown together
with the difference $\Delta$$\equiv$$\delta E - \delta E_{\rm sc}$. All
energies are in MeV.}
\begin{tabular}{ccccccc}
\multicolumn{7}{c}{Neutrons}\\
 Nucleus & $\delta E$ & $\sigma$ & $\tilde{\lambda}$ & $\delta E_{\rm sc}$
  & $\lambda_{\rm sc}$ & $\Delta$ \\
\hline
$^{78}$Ni &   -2.83& 0.183& -2.64 & -4.22& -2.51 & 1.39\\
$^{90}$Zr &   -7.19& 0.100& -9.63 & -6.82& -9.77 &-0.37\\
$^{96}$Zr &    0.24& 0.016& -7.32 & 0.82& -7.37 & -0.58\\
$^{104}$Zr &   6.57& 0.056& -4.79 & 6.48& -4.71 & 0.09\\
$^{106}$Zr &   5.97& 0.039& -4.23 & 5.56& -4.13 & 0.41\\
$^{108}$Zr &   5.76& 0.150& -3.69 & 4.94& -3.57 & 0.82\\
$^{110}$Zr &   4.49& 0.029& -3.17 & 3.45& -3.05 & 1.04\\
$^{122}$Zr &  -4.61& 0.056& -0.32 & -6.40& -0.44 & 1.79\\
$^{124}$Zr &  -2.91& 0.052&  0.12 &  -4.39& -0.12 & 1.47\\
$^{132}$Sn &  -8.70& 0.023& -4.50 & -8.94& -4.42 & 0.24\\
$^{146}$Gd &  -10.09& 0.118& -9.77 & -9.85& -9.89 & -0.24\\
$^{208}$Pb &  -11.37& 0.063& -5.50 & -11.23& -5.56 & -0.13\\
$^{298}$114 &  -8.44& 0.090& -4.83 & -8.63& -4.81 &0.19\\[2mm]
\multicolumn{7}{c}{Protons}\\
 Nucleus & $\delta E$ & $\sigma$ & $\tilde{\lambda}$ & $\delta E_{\rm sc}$
  & $\lambda_{\rm sc}$ & $\Delta$ \\
\hline
$^{48}$Ni&  -2.11& 0.084& -0.16 & -1.94& -0.08 & -0.17\\
$^{90}$Zr&  1.96& 0.222& -6.65 &1.45& -6.84 &0.51\\
$^{100}$Sn&  -7.31& 0.083& 0.72 & -7.01& 0.61 & -0.30\\
$^{132}$Sn&  -6.02& 0.081& -13.24 & -6.65& -13.31 & 0.63\\
$^{146}$Gd&  5.27& 0.247& -3.98 & 4.51& -4.12 &0.77\\
$^{180}$Pb&  -7.72& 0.016& -0.81 & -8.57& -0.87 &0.85\\
$^{208}$Pb&  -6.73& 0.028& -7.16 & -7.33& -7.22 &0.60\\
\end{tabular}
\label{tab1}
\end{table}

\end{document}